\documentclass{article}

\usepackage{epsfig}

\author{Arindam RoyChoudhury} 

\def\fnr{f_{\mathbf{N,R}}}
\def\fstr{f_{\mathbf{N,R}}^{\ast}}
\def\lp{\Lambda^{(P)}}
\def\lst{\Lambda^{(P')\ast}}
\def\ltwo{\Lambda^{(2)}}
\def\l4{\Lambda^{(4)}}

\def\na{n_A}
\def\nb{n_B}
\def\nab{n_{AB}}
\def\nd{n_D}
\def\ia{z_A}
\def\ib{z_B}
\def\iab{z_{AB}}
\def\id{z_D}
\def\ta{\tau_A}
\def\tb{\tau_B}
\def\tab{\tau_{AB}}
\def\td{\tau_D}
\def\ra{r_A}
\def\rb{r_B}
\def\rab{r_{AB}}
\def\rd{r_D}
\def\b1{(1)}
\def\b2{(2)}
\def\bft{\mathbf{\Psi}}
\def\prn{\mbox{Pr}_{\mathbf{n}}}
\def\p2r{\mbox{Pr}}
\def\bfn{\mathbf{n}}
\def\btheta{\mbox{\boldmath$\theta$\unboldmath}}

\def\a0{\mbox{A}(0)}
\def\b0{\mbox{B}(0)}
\def\c0{\mbox{C}(0)}
\def\g0{\mbox{G}(0)}

\def\s0b{s_0^{(\mbox{\scriptsize{b}})}}

\def\m0b{m_0^{(\mbox{\scriptsize{b}})}}

\begin{document}

\title{Identifiability of a Coalescent-based Population Tree Model} 



\maketitle

\begin{abstract}
Identifiability of evolutionary tree models has been a recent topic of discussion and some models have been shown to be non-identifiable.
A coalescent-based rooted population tree model, originally proposed by Nielsen et al. 1998 \cite{nea98}, has been used by many authors in the 
last few years and is a simple tool to accurately model the changes in allele frequencies in the tree.
However, the identifiability of this model has never been proven. Here we prove this model to be identifiable by showing that the model
parameters can be expressed as functions of the probability distributions of subsamples. This a step toward proving the consistency of the maximum likelihood
estimator of the population tree based on this model.
\end{abstract}

\section{Introduction}

A rooted evolutionary tree is a directed weighted tree graph; it represents the evolutionary relationship between groups 
(also called taxa) of organisms (Figure 1(a)). A leaf or a tip is a node with degree 1;
each tip represents a modern day taxon. The root (node 0) represents the most recent common ancestor (MRCA) of all the taxa. 
The direction (of evolution) is from the root to the tips. Evolutionary tree as a vector of parameters influences the probability
distribution of alleles at the tips.

\begin{figure}
\includegraphics{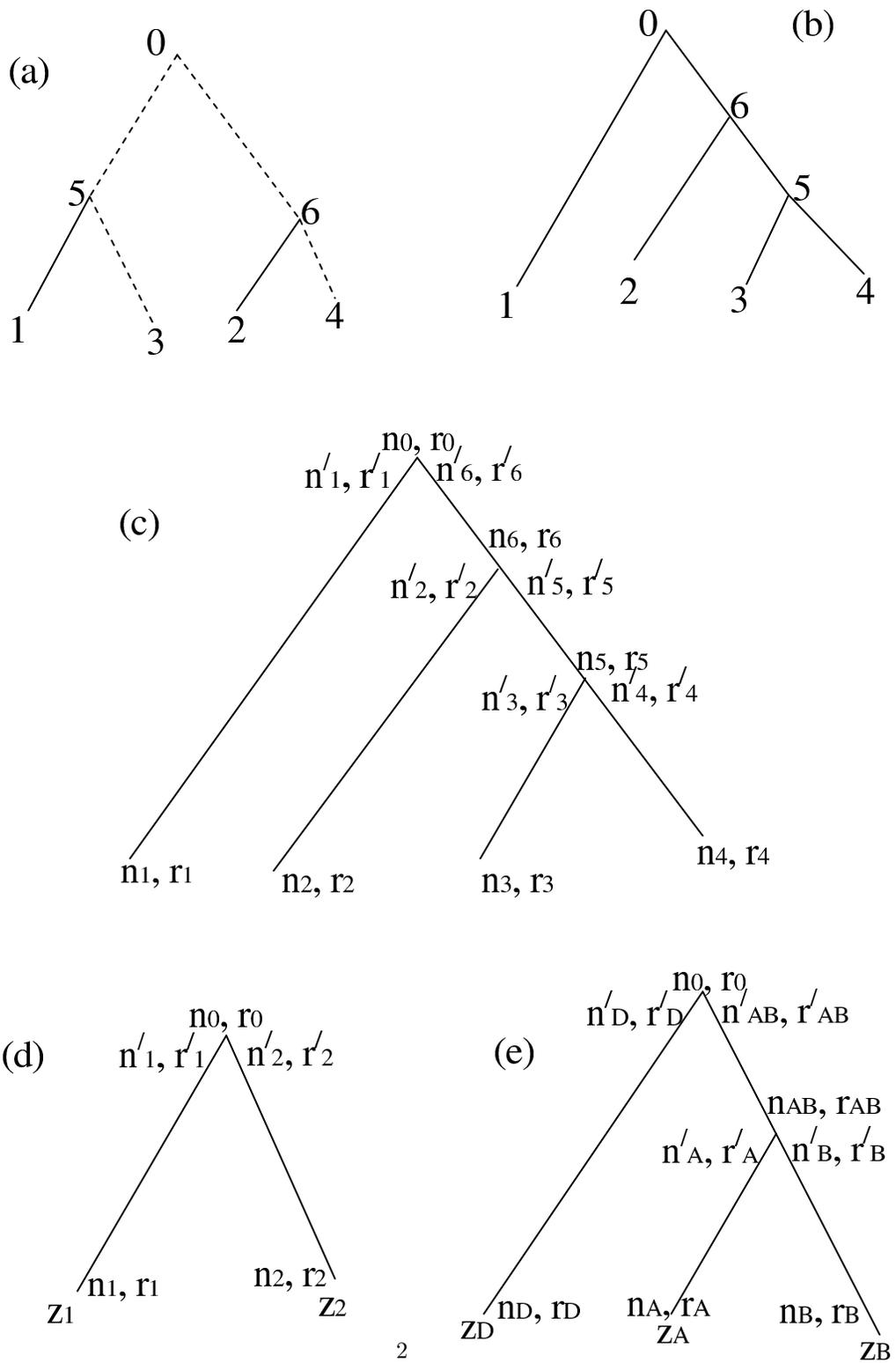}
\caption{Population trees}
\label{fig1}
\end{figure}

A rooted population tree is a rooted evolutionary tree where the taxa are populations from the same species. 
Two types of parameters are common in any model of the rooted population tree: the tree-topology parameter (a categorical parameter) 
for the whole tree, 
and a branch parameter for each branch (also called edge). 

The tree-topology is the order in which the path from the root separates for the given set of populations; it is represented
as a directed tree graph without the weight.
(In Figure 1(a) and (b), the two trees have different tree-topologies for the populations 1-4.) 
A branch parameter is usually a branch-length (an edge-weight) 
or a transition probability matrix that influences the change in allele frequency between the two nodes of a branch.

Here we will prove the identifiability of a population tree model by \cite{nea98,rea08} that uses Kingman's Coalescent Process (\cite{k82a}). The model was later modified 
and expanded by various authors (\cite{ns00,rea08,r11,rt12,bea12}). Coalescent-based models are of significant importance as they model the underlying allele frequency
changes with accuracy and relative ease (see \cite{lea09b}). 

Due to the underlying structure in evolutionary tree-based models, its identifiability is never obvious. The identifiability of certain 
evolutionary tree models have been a recent topic of discussion. \cite{aea08} proved the identifiability of a general time reversible (GTR) 
transition probability matrix-based model. Non-identifiability of another time reversible model was established in \cite{sea94}. The non-identifiability of 
mixture models have been discussed in \cite{ms07}.
The identifiability for the \cite{f81} model has been proven by \cite{ch10}. To our knowledge the identifiability of the 
coalescent-based model of \cite{nea98,rea08} has never been proven.

For estimating evolutionary trees each independent genetic locus is viewed as a single data-point, 
as opposed to viewing each individual as a data-point (see, for example \cite{r11}). Thus, 
identifiability would mean that the model parameters can be identified from the distribution of allele-types for a set of individuals
at a single genetic locus.

\section{The model}

\label{sec:model}

In this section we will describe the underlying model of \cite{nea98,rea08}. We start by defining our notations (see also Figure 1(c)).
We define a $P$-tip population tree as $T = (\lp,\bft,\btheta)$. 
The parameter $\lp$ is the tree-topology, an unweighted directed tree-graph; it takes finitely many discrete categorical values; the $(P)$ in superscript denotes
the number of tips. 
The parameter 
$\bft = (\tau_1,\tau_2,\dots,\tau_{2P-2})$ is a vector of length $2P-2$ consisting of the
branch-lengths $\tau_i$ for each branch $i$ in $\lp$. 
A strictly bifurcating tree-topology has exactly $2P-2$ branches. If $\lp$ is non-bifurcating then it has less branches
and the remaining elements of $\bft$ are populated by zeros.
The parameter $\btheta$ is a vector containing the parameters of root distribution
which we will define later in this section. We also define $S(\lp)$ as the set of tips at $\lp$.

At each tip $z$ there are $n_z (\geq 1)$ lineages, each having allele-type `0' or `1'. The allele types among these lineages at each tip are the observable random variables.
Similarly, at each non-tip node $x$, the random variable $n_x (\geq 1)$ is the (random) number of lineages that are ancestral to the tips below $x$ along the tree. We also
define the random variable $r_x$ at each node $x$ (tip or non-tip), as the count of allele `1' among the $n_x$ lineages. From now
on we will use the term `allele-count' to refer to the count of allele `1'. For each tip $z$, the allele-count $r_z$ is observable.

Consider a branch with lower (towards the tips) node $x$ and upper (towards the root) node $y$. Let $n_x'$ be the number of lineages in $y$ that are ancestral to the $n_x$ lineages
at $x$ ($n_x' \leq n_x$).
Also, let $r_x'$ be the allele-count among these $n_x'$ lineages ($r_x' \leq r_x$). If $y$ is the upper node of $\nu$ branches with lower nodes
$x_1, x_2, \dots, x_{\nu}$, then
\begin{eqnarray}
n_{x_1}, n_{x_2}, \dots, n_{x_{\nu}} \mbox{ are independent, and } n_y = \sum_{k=1}^{\nu} n_{x_k}' \label{eq:nsum}
\end{eqnarray}
and also $r_y = \sum_{k=1}^{\nu} r_{x_k}'$.
(For a strictly bifurcating tree $\nu = 2$.)

From the model parameters $T = (\lp,\bft,\btheta)$ one computes the probability of observed vector of allele-counts 
$\mathbf{r} = (r_1, r_2, \dots, r_P)$ from samples of sizes $\bfn = (n_1, n_2, \dots, n_P)$ at $P$ tips ($1,2,\dots,P$)
as follows. 
Consider a branch with length $\tau_{x_1}$, with upper node $y$ and lower node $x_1$. Given
the probability mass function (pmf) of $n_{x_1}$ (the number of lineages at $x_1$), the pmf of $n_{x_1}'$ 
is computed as
\begin{eqnarray}
\prn (n_{x_1}' = i'  \, \vert \, n_{x_1}  = i; \tau_{x_1}) 
= \biggl(\prod_{j=i'+1}^{i} \lambda_j\biggr) \sum_{j=i'}^{i} {e^{-\lambda_j\tau_{x_1}}
\over \prod_{j' = i', j' \neq j}^i (\lambda_{j'} - \lambda_j)}, \label{eq:tn85}
\end{eqnarray}
where $\lambda_j = j(j-1)/2$. Then, the pmf of $n_y$ is determined from Eq. (\ref{eq:nsum}).

Using Eqs. (\ref{eq:tn85}) and (\ref{eq:nsum}), starting from $\bfn = (n_1, n_2, \dots, n_P)$ and going upward, one 
computes the pmf of $n_z$ and $n_z'$ for any non-tip non-root node $z$, and finally $n_0$ at the root (node 0).
Then a `root distribution' with parameter $\btheta$ gives the pmf of (allele-count) $r_0$ given $n_0$ at the root:
$$\g0 = \left(\prn (r_0 = j \vert n_0 = i; \btheta),j=0,1,\dots,n_0; i=1,2,\dots, \m0b  \right),$$
where 
$$\m0b = \sum_{z \mbox{ \scriptsize{is a tip}}} n_z$$
is the maximum possible value of $n_0$ (number of lineages at the root).
Different authors have used different root distributions. In particular \cite{rea08} used symmetric Beta-Binomial distribution: 
\begin{eqnarray}
\prn (r_0 = j \vert n_0 = i; \theta) = {i \choose j} \frac{\beta(j+\theta)\,\beta(i-j+\theta)}{\beta(\theta,\theta)}, \label{eq:betabinom}
\end{eqnarray}
where $\beta(.,.)$ is the Beta Function; $\theta > 0$ is a parameter to be estimated.

Then, from the distribution of $n_0, r_0$ and $(n_z,n_z')$ for all non-root nodes $z$, we compute the distribution of $r_z$ (allele-counts) at the rest of the nodes as 
follows.
Consider a node $y$ where
$\nu$ branches merge from the bottom with the bottom nodes $x_1,x_2,\dots,x_{\nu}$.
Recall that we already have the distributions of $n_y$, $n_{x_i}$ and $n_{x_i}'$,
$i = 1,2,\dots,\nu$. The pmf of $r_{x_i}'$ is computed from the pmf of $r_y$ using
the formula
\begin{eqnarray}
& & \prn (r_{x_1}' = j_1',r_{x_2}' = j_2',\dots,r_{x_{\nu}}' = j_{\nu}' \, \vert \, r_y = j, n_y = i, n_{x_1}' = i_1',n_{x_2}' = i_2',\dots,n_{x_{\nu}}' = i_{\nu}' ) \nonumber \\
& = & \frac{{j \choose j_1',j_2',\dots,j_{\nu}'} \, {i-j \choose i_1'-j_1',i_2'-j_2',\dots,i_{\nu}'-j_{\nu}'}}{ {i \choose i_1',i_2',\dots,i_{\nu}'}}. \label{eq:hpgeo}
\end{eqnarray}
Then the pmf of $r_{x_k}$ is computed from the above pmf using the following (from an expression in \cite{rea08}):
\begin{eqnarray}
& & \prn (r_{x_k} = j_k \, \vert \, r_{x_k}' = j_k', n_{x_k}' = i_k', n_{x_k} = i_k) \nonumber \\
& = & 
\frac{\beta(j_k,i_k-j_k)}{\beta(j_k',i_k'-j_k')}{i_k - i_k' \choose j_k - j_k'}, 
0 < j_k < i_k \mbox{ and } 0 < j_k' < i_k', \cr 
                                                                                     1, 
0 = j_k = j_k' \mbox{ or } 0 = i_k - j_k = i_k' - j_k', \cr 
                                                                                     0, 
\mbox{ otherwise;}
\label{eq:polya}
\end{eqnarray}
$k=1,2,\dots,\nu$ (\cite{rea08}). Thus, starting with $\g0$ at the root, 
one computes the joint pmf of $(r_1,r_2,\dots,r_P)$ from the formulae
in Eqs. (\ref{eq:hpgeo}) and (\ref{eq:polya}).
Note that in Eqs. (\ref{eq:nsum}), (\ref{eq:tn85}), (\ref{eq:betabinom}), (\ref{eq:hpgeo}) and (\ref{eq:polya})
probability `flows' up along $n$'s and then flows down along $r$'s. 

Now that we have completely described the model, we will proceed to prove the identifiability of this model in the next section.

\section{Identifiability}

Let $T = (\lp,\bft,\btheta)$ be a tree with $S(\lp) = \{1,2,\dots,P\}$. We define a subtree 
$T^{\ast}$ of $T$ as a tree formed by a subset $S^{\ast}$ (cardinality $P' \leq P$) of $S(\lp)$ by tracking the tips 
in $S^{\ast}$ along the tree to their most recent common ancestor (MRCA) node. 
Thus, $T^{\ast} = (\lst,\bft^{\ast},\btheta)$, where $\lst$ is the tree-topology with $P'$ tips of $S^{\ast}$.
For example, 
in Figure 1(a), $P=4$, $S(\l4)= \{1,2,3,4\}$, $S^{\ast} = \{3,4\}$ and the subtree 
$T^{\ast}$ is drawn with the dotted lines.

Consider two distinct trees $T_1 = (\lp_1,\bft_1,\btheta_1)$ and $T_2 = (\lp_2,\bft_2,\btheta_2)$ with a common set of tips 
$S_{T_{1,2}} = S(\lp_1) = S(\lp_2)$.

If $\btheta_1 = \btheta_2 = \btheta$, then there must be at least one doubleton subset $\{z_1,z_2\} \subseteq S_{T_{1,2}}$ with the following property: 
the subtrees 
$T_1^{\ast}=( \ltwo ,\bft^{\ast}_1,\btheta)$ and $T_2^{\ast}=( \ltwo ,\bft^{\ast}_2,\btheta)$, formed by tracking $z_1$ and $z_2$ to the root
in $T_1$ and $T_2$ (respectively), 
are distinct. 
That is, if $\bft^{\ast}_l = (\tau_{1l},\tau_{2l})$
and $\tau_{jl}$ is the path distance (total branch length) between $z_j$ and the MRCA of $z_1$ and $z_2$ along the subtree 
$T_l^{\ast}$ ($j,l=1,2$), then 
$(\tau_{11},\tau_{21}) \neq (\tau_{12},\tau_{22}).$
(Note that there is only one possible tree-topology for a two-tip tree, denoted as $ \ltwo $ above.) 
Thus, the set of all two tip subtrees, along with $\btheta$, uniquely identifies the tree.

We assign the two-tip subtrees into two categories: Type-I subtrees are those with the root as the 
MRCA of the two tips. For example in Figure 1(a), the subtree formed by tips $\{3,4\}$ has the root as 
the MRCA of the two tips 3 and 4. Thus, it is of Type-I. All other two-tip subtrees are Type-II subtrees. 
For example, in Figure 1(a), if a subtree is formed by tips 2 and 4, it will be a Type-II subtree as their 
MRCA is node 6, and not the root. We will deal with these two types of subtrees separately.

We note that the root distribution of \cite{rea08} (Eq. (\ref{eq:betabinom})) is identifiable as it is Beta-Binomial. 
Next, we will prove the identifiability of the whole model by assuming a general identifiable root distribution that has 
parameter vector $\btheta$. (In particular, our proof would work with Beta-Binomial
as the root distribution.)


\textbf{Theorem}
Suppose that we have a tree $T$ with the underlying model as described in Section \ref{sec:model}. Also, suppose that
we have $N_k \geq 2$ lineages sampled at each tip $k$ and the root distribution is identifiable. Then the parameters of $T$ are identifiable from
the distribution of allele types at the tips. 

To prove the above theorem, we will show that the parameters of each two-tip subtree can be expressed as a function of the 
joint pmf 
\begin{eqnarray}
\left(\prn \left((R_1, R_2, \dots, R_P) = (J_1,J_2,\dots,J_P); T \right), J_k=0,1,2,\dots,N_k, k=1,2,\dots,P \right). \nonumber \\
\label{eq:probd}
\end{eqnarray}
This will complete the proof as the set of all two-tip 
subtrees, along with $\btheta$, uniquely identifies the tree.

\subsection{Identifiability of Type-I subtrees}

\label{subsec:type1}

Suppose that $T = (\ltwo ,\{\tau_1,\tau_2\},\btheta)$ is a Type-I subtree with the underlying model as described in Section \ref{sec:model}. 
Let $z_1$ and $z_2$ be its two tips.
Let the root be denoted as `0' (Figure 1(d)) and let $\tau_k$ be path distance between $z_k$ and the root $(k=1,2).$

\textbf{Proposition}
Suppose that we have at least
two lineages sampled at each of $z_1$ and $z_2$ and the root distribution is identifiable. Then $\tau_1, \tau_2$
and $\btheta$ can be expressed as functions of 
the joint pmf of allele types in $z_1$ and $z_2$, and hence they are identifiable.
\label{prop:type1}

\textbf{Proof}
Suppose that we have samples of $N_1$ and $N_2$ lineages from $z_1$ and $z_2$ respectively, and the allele-counts among these lineages are $R_1$ and $R_2$ respectively.
Let the joint pmf of $(R_1,R_2)$ be $\fnr$.

Consider random subsamples (without replacement) of size $n_1$ and $n_2$ from $z_1$ and $z_2$ respectively with $n_k \leq 2, k=1,2$. 
Rather than working with the allele-counts $R_k$ at the original samples, we will work with allele-counts $r_k$ at the subsamples.

One computes the joint pmf of $(r_1,r_2)$ from $\fnr$ as
\begin{eqnarray}
& & \prn \left(r_k=j_k,\,k=1,2 \, \vert \, R_k = J_k, N_k = I_k, n_k = i_k,\,k=1,2;\tau_1, \tau_2, \btheta \right) \nonumber \\
& = & \sum_{J_1 = j_1}^{I_1 - (i_1 - j_1)} \sum_{J_2 = j_2}^{I_2 - (i_2 - j_2)}
\left( \prod_{k=1}^2 \frac{{J_k \choose j_k} \, {I_k - J_k \choose i_k - j_k}}{{I_k \choose i_k}} \right) \, \fnr (J_1,J_2).
\nonumber
\end{eqnarray}
We will argue that the joint pmfs
$(r_1,r_2)$ for $(n_1,n_2) = $ (1,1), (1,2) and (2,1)
are enough to identify the parameters $\tau_1, \tau_2$ and $\btheta$. 

As before, let $n_k'$ be the number of lineages ancestral to subsamples at $z_k$ that are
present at the top node (the root) (see Figure 1(d)) and $r_k'$ be the allele-count out of these $n_k'$; ($k=1,2$). 
Also, let $n_0 = n_1' + n_2'$ be the number of lineages at the root ancestral 
to the subsampled lineages at $z_1$ and $z_2$, and $r_0 = r_1' + r_2'$ be the allele-count out of these $n_0$ lineages. 

First, consider the case $n_1 = n_2 = 1$. Then $r_k = $ 0 or 1 for $k=1,2$. 
From Eq. (\ref{eq:tn85}) it follows that $n_1' = n_2' = 1$; thus, $\prn (n_k' = i'  \, \vert \, n_k  = i; \tau_k)$ and hence
$\p2r(r_1 = j_1, r_2 = j_2\,\vert\, n_1=n_2=1;\tau_1,\tau_2,\btheta)$
does not involve $\tau_1$ and $\tau_2$. 
From Eq. (\ref{eq:polya}) it also follows that $r_k = r_k', k=1,2$. Also, $n_0 = n_1' + n_2' = 1 +  1 = 2$.

Note that $r_0 = r_1' + r_2'$ and $r_k = r_k'$ ($k=1,2$) are counts.
Thus,
$$(r_1,r_2) = (0,0) \Longleftrightarrow (r_1',r_2') = (0,0) \Longleftrightarrow r_0 = 0.$$
Using a symmetric argument
$$(r_1,r_2) = (1,1) \Longleftrightarrow (r_1',r_2') = (1,1) \Longleftrightarrow r_0 = 2.$$
Thus,
\begin{eqnarray}
\p2r \left(\,(r_1,r_2) = (j,j) \, \vert \, n_1 = n_2 = 1; \btheta \right) = \p2r \left(r_0 = 2\,j \, \vert \, n_0 = 2; \btheta \right), \hspace{0.2 in} j=0,1.
\label{eq:rootf1}
\end{eqnarray}
It follows that
\begin{eqnarray}
& & \p2r \left(\,(r_1,r_2) = (0,1)\, \vert \, n_1 = n_2 = 1 ;\btheta \right) + \p2r \left(\,(r_1,r_2) = (1,0)\, \vert \, n_1 = n_2 = 1 ;\btheta \right) \nonumber \\
& = & 1 - \p2r \left(\,(r_1,r_2) = (0,0) \, \vert \, n_1 = n_2 = 1; \btheta \right) 
- \p2r \left(\,(r_1,r_2) = (1,1)\, \vert \, n_1 = n_2 = 1;  \btheta \right) \nonumber \\
& = & 1 - \p2r \left(r_0 = 0\, \vert \, n_0 = 2; \btheta \right) - \p2r \left(r_0 = 2\, \vert \, n_0 = 2; \btheta \right) \nonumber \\
& = & \p2r \left(r_0 = 1\, \vert \, n_0 = 2; \btheta \right)
\label{eq:rootf2}
\end{eqnarray}
Thus, from Eqs. (\ref{eq:rootf1}) and (\ref{eq:rootf2}) $\p2r \left(r_0 = j_0 \,\vert\, n_0 = 2; \btheta \right), \, j_0 = 0,1,2$ can be expressed as functions
of $\p2r \left(\,(r_1,r_2) = (j_1,j_2) \, \vert \, n_1 = n_2 = 1; \btheta \right), \, j_1,j_2 = 0,1$. 
The former is the root distribution for $n_0 = 2$, which is identifiable 
by the condition of Proposition \ref{prop:type1}.
Thus, $\btheta$ can the expressed as a function of the pmf of $r_0$ (given $n_0 = 2$), and thus as a function of joint pmf of $(r_1,r_2)$. 
Hence, it can also be expressed as a function of $\fnr$.

Next, we consider $n_1 = 2, n_2 = 1$. Then $r_1 = $ 0, 1 or 2 and $r_2 = $ 0 or 1.
From Eq. (\ref{eq:tn85}) it follows that $n_2' = 1$; thus $\prn (n_2' = i_2'  \, \vert \, n_2  = i_2; \tau_2)$ and hence
$\p2r((r_1,r_2) = (0,1)\,\vert\, n_1=n_2=1;\tau_1,\tau_2,\btheta)$
does not involve $\tau_2$. 
Moreover, $n_0 = n_1' + n_2' = n_1' + 1$. 
Also, from Eq. (\ref{eq:polya}) it follows that
$$(r_1,r_2) = (0,1) \Longleftrightarrow (r_1',r_2') = (0,1).$$
Thus,
\begin{eqnarray}
& & \p2r \left(\,(r_1,r_2) = (0,1)\,\vert\,(n_1,n_2)=(2,1); \tau_1, \btheta \right) \nonumber \\
& = & \sum_{i'=1}^2 \p2r \left(\,(r_1',r_2') = (0,1) \, \vert \, (n_1',n_2')=(i',1);\btheta \right) \, 
\p2r \left(n_1'=i'\,\vert\,n_1=2; \tau_1 \right) \nonumber \\
& = & \sum_{i'=1}^2 \p2r \left(\,(r_1',r_2') = (0,1) \, \vert \, n_0=n_1'+1=i'+1;\btheta \right) \, \p2r \left(n_1'=i'\,\vert\,n_1=2; \tau_1 \right) \nonumber \\
& = & \sum_{i'=1}^2 \sum_{j_0=0}^{i'+1} \p2r \left(\,(r_1',r_2') = (0,1) \, \vert \, r_0 = j_0, n_0=i'+1 \right) \nonumber \\
& & \, \times \, \p2r \left(r_0 = j_0 \, \vert \, n_0=i'+1;\btheta \right)
\, \p2r \left(n_1'=i' \,\vert \,n_1=2;\tau_1 \right) \nonumber
\end{eqnarray}
Note that $r_0 \neq 1 \Longrightarrow (r_1',r_2') \neq (0,1)$. Also, note that 
$$ \p2r \left(r_0 = 1 \, \vert \, n_0=i'+1;\btheta \right) $$
is a function of $\btheta$ only (and no other parameters); hence we call it $c_{i'+1}(\btheta),\, i'=1,2$. 
Thus,
\begin{eqnarray}
& & \p2r \left(\,(r_1,r_2) = (0,1)\,\vert\,(n_1,n_2)=(2,1); \tau_1, \btheta \right) \nonumber \\
& = & \sum_{i'=1}^2 \p2r \left(\,(r_1',r_2') = (0,1) \, \vert \, r_0 = 1,\, n_0=i'+1 \right) 
\, c_{i'+1}(\btheta) \, \p2r \left(n_1'=i' \,\vert \,n_1=2;\tau_1 \right) \nonumber \\
& = & \frac{c_2(\btheta)}2 \, (1-e^{-\tau_1}) + \frac{c_3(\btheta)}3 \, e^{-\tau_1}
  = e^{-\tau_1} \, \left(\frac{c_3(\btheta)}3 - \frac{c_2(\btheta)}2 \right) + \frac{c_2(\btheta)}2 \nonumber
\end{eqnarray}
from Eqs. (\ref{eq:tn85})  and (\ref{eq:hpgeo}).
From the above equation it follows that
\begin{eqnarray}
\tau_1 = b \, \Bigl( \, \p2r \left(\,(r_1,r_2) = (0,1) \, \vert \, (n_1,n_2) = (2,1); \tau_1, \btheta \right), \, \btheta \Bigr)
\label{eq:tau1}
\end{eqnarray}
for some function $b(.,.)$.
We have already established that $\btheta$ can be expressed as a function of $\fnr$.
Thus, $\tau_1$ can be expressed as a function of $\fnr$ and hence $\tau_1$ is identifiable.

Using a symmetric argument, one can establish that $\tau_2$ can be expressed as a function of $\fnr$ and hence it is identifiable. Thus, this proposition is proven.

\subsection{Identifiability of Type-II subtrees}

Consider a Type-II subtree of with tips $\ia$ and $\ib$. Let the MRCA node of $\ia$ and $\ib$ be denoted as $\iab$. (By definition
$\iab$ is not the root.) Also, consider the path from $\iab$ to the root (node 0) and call it branch $AB$. There must be at least another branch $H$ attached to the
root other than branch $AB$ (Figure 1(e)). Consider a tip $\id$, such that the path between $\id$ and the root goes through $H$. Let $\ta$ be the path
distance between $\iab$ and $\ia$ and let $\tb$ be the path
distance between $\iab$ and $\ib$. Also, let $\tab$ be the path distance between the root and $\iab$ and let $\td$ be the path distance between the root and $\id$.

\textbf{Proposition}
Suppose that we have at least
two haploids sampled at each of $\ia, \ib$ and $\id$ and the root distribution is identifiable. Then $\ta, \tb, \tab, \td$ and $\btheta$
can be expressed as functions of the joint pmf of the allele types at $\ia, \ib$ and $\id$, and hence they are identifiable.

\textbf{proof}
Suppose that we have samples of $N_A$, $N_B$ and $N_D$ lineages from $\ia$, $\ib$ and $\id$ respectively, and the allele-counts among these lineages are 
$R_A$, $R_B$ and $R_D$ respectively.
Let the joint pmf of $(R_A,R_B,R_D)$ be $\fstr$.

First we consider the Type-I subtree formed by $\ia$ and $\id$. From Proposition \ref{prop:type1} one can establish that $\btheta$, $\td$ and 
$\ta + \tab$ can be expressed as a function of the joint pmf of $(R_A,R_D)$ and hence of $\fstr$. 
A symmetric argument also establishes that $\tb + \tab$ 
can be expressed as functions of $\fstr$.
Next we will show that each of $\ia, \ib$ and $\iab$ can be expressed as function of $\fstr$.

Consider a random subsample of size one from each of $\ia, \ib$ and $\id$. Let $\na, \nb$ and $\nd$ be the numbers of subsampled haploids at $\ia, \ib$ and $\id$ respectively. 
(Thus, $\na=\nb=\nd=1$).
Let $\ra, \rb$ and $\rd$, respectively, be the observed allele-counts
at these subsamples. ($r_k = $ 0 or 1 for $k=A,B,D$.)
As before, let $n_k'$ be the number of lineages ancestral to subsamples at $z_k$ that are
present at the top node of the branch (in the subtree) attached to $z_k$ (see Figure 1(e)) and $r_k'$ be the allele-count out of these $n_k'$ ($k=A,B,D$). 

From Eq. (\ref{eq:tn85}) it follows that
$\na = \nb = \nd = n_A' = n_B' = n_D' = 1$
and thus
$ \p2r \left(n_k' = i_k' \, \vert \, n_k = i_k; \tau_k \right)$
does not involve $\tau_k$ ($k=A,B,D$). Hence,
\begin{eqnarray} 
\p2r \left( (\ra,\rb,\rd) = (0,0,1) \, \vert \, \na=\nb=\nd=1; \ta, \tb, \tab, \td, \btheta \right) \nonumber
\end{eqnarray}
does not involve $\ta, \tb$ and $\td$.
Also,
\begin{eqnarray} 
& & \p2r \left( (\ra,\rb,\rd) = (0,0,1) \, \vert \, \na=\nb=\nd=1; \tab, \btheta \right) \nonumber \\
& = & \sum_{J_A = j_A}^{I_A - (i_A - j_A)} \sum_{J_B = j_B}^{I_B - (i_B - j_B)} \sum_{J_C = j_C}^{I_C - (i_C - j_C)}
\left( \prod_{k \in \{A,B,D\}} \frac{{J_k \choose j_k} \, {I_k - J_k \choose i_k - j_k}}{{I_k \choose i_k}} \right) \, \fstr (J_A,J_B,J_D). \nonumber \\
\label{eq:rabd}
\end{eqnarray}
Thus, the left side of Eq. (\ref{eq:rabd}) can be expressed as a function of $\fstr$.
It also follows from Eq. (\ref{eq:polya}) that $r_k = r_k', k=A,B,D$.

Let $\nab = n_A' + n_B'$ be the total number of lineages from subsamples of $\ia$ and $\ib$ that are present at node $AB$, and let $\rab = r_A' + r_B'$ be the allele-counts
out of these $\nab$ lineages. Also, let $n_{AB}'$ be the number of lineages ancestral to those $\nab$ lineages that are
present at the top node (root) of the branch $AB$, and let $r_{AB}'$ be the allele-count out of these $n_{AB}'$ lineages. As before, let $n_0 = n_{AB}' + n_D'$ be
the total number of lineages at the root ancestral to the subsamples at $\ia, \ib$ and $\id$; let $r_0 = r_{AB}' + r_D'$ be the allele-count out of these $n_0$ lineages. Note that 
$\nab= n_A' + n_B' = 2,$ $\, n_{AB}' \leq \nab$.
From Eq. (\ref{eq:polya}) and the fact that $\rab = r_A' + r_B'$ it follows that
$$(\ra,\rb) = (0,0) \Longleftrightarrow (r_A',r_B') = (0,0) \Longleftrightarrow \rab = 0.$$
Thus,
\begin{eqnarray}
& & \p2r \left( (\ra,\rb,\rd) = (0,0,1) \, \vert \, \na=\nb=\nd=1; \tab, \btheta \right) \nonumber \\
& = & \p2r (\,(\rab,\rd) = (0,1) \, \vert \, (\nab,\nd)=(2,1); \tab, \btheta) \label{eq:r_ab}
\end{eqnarray}
Consider the part of the subtree consisting of the path from $\iab$ and $\id$ to the root; it is a Type-I subtree with $\iab$ and $\id$ as the tips, 
and $\tab$ and $\td$, respectively, as the lengths of the attached branches; it has $(\nab,\nd) = (2,0)$, respectively, as the numbers of observed lineages
at $\iab$ and $\id$
and $(\rab,\rd)$, respectively, 
as the allele-counts in these lineages. From Eq. (\ref{eq:tau1}) and (\ref{eq:r_ab})
\begin{eqnarray}
\tab 
& = & b \, \Bigl( \, \p2r \left(\,(\rab,\rd) = (0,1) \, \vert \, (\nab,\nd)=(2,1); \tab, \btheta \right), \, \btheta \Bigr) \nonumber \\
& = & b \, \Bigl( \, \p2r \left( (\ra,\rb,\rd) = (0,0,1) \, \vert \, \na=\nb=\nd=1; \tab, \btheta \right), \, \btheta \Bigr). \nonumber
\end{eqnarray}
As we have already established that $\ta + \tab$, $\tb + \tab$, $\td$, $\btheta$ and the left side of Eq. (\ref{eq:rabd}) 
can be expressed as functions of $\fstr$,
it follows that $\ta, \tb, \tab, \td$ and $\btheta$ can be expressed as 
functions of $\fstr$. Thus, they are identifiable and this proposition is proven.

Thus, the parameters of the tree are identifiable, as each two-tip subtree along with the root distribution parameter $\btheta$ is identifiable.

\section{Discussions}

We have proven that the model parameters are identifiable under the coalescent-based population tree model of \cite{nea98,rea08}. 
Thus, the problem of estimation of population tree from this model is indeed meaningfully stated. Moreover, as 
identifiability is a required condition for consistency of maximum likelihood estimator (MLE), this is a step towards proving the
consistency of MLE for this model. We have proven the identifiability of the tree parameters for any identifiable root distribution. As a result
our proof is valid for different versions of this model (that vary at the root distribution) such as \cite{nea98,rea08,r11}.

\end{document}